\documentclass[useAMS,usenatbib]{mn2e}
\usepackage{graphicx}
\usepackage{amssymb}
\usepackage{color}

\title[The contribution from blazar cascade emission to the EGRB]{The contribution from blazar cascade emission
to the extragalactic gamma-ray background: What a role does the
extragalactic magnetic field play ?}
\author[Yan et al.]{Dahai Yan, Houdun Zeng, and Li Zhang\thanks{E-mail: lizhang@ynu.edu.cn}\\
Department of Physics, Yunnan University, Kunming, China}

\begin{document}
\pagerange{\pageref{firstpage}--\pageref{lastpage}} \pubyear{2010}

\maketitle

\label{firstpage}

\begin{abstract}
We estimate the contribution to the extragalactic gamma-ray
background (EGRB) from both intrinsic and cascade emissions
produced by blazars using a simple semi-analysis method for two
models of the blazar gamma-ray luminosity function (GLF). For the
cascade emission, we consider two possible contributions: one is
due to that the flux of the cascade emission is lower than the
{\it Fermi} LAT sensitivity (case I), which is independent on the
extragalactic magnetic field (EGMF), another is due to the fact
that the flux of the cascade emission is larger than the {\it
Fermi} LAT sensitivity but the emission angle is larger than LAT
point-spread-function (PSF) angle (case II), which depends on the
EGMF. Our results indicate that (1) blazar contribution to the
EGRB is dominant although it depends on the GLF model and the
EGMF; (2) the EGMF plays an important role in estimating the
contribution from the cascade emission produced by blazars, the
contribution from the cascade emission will significantly alter
the EGRB spectrum when the strength of the EGMF is large enough (
say $B_{\rm EGMF}>10^{-12}\ $G); and (3) since the cascade
emission in case II reaches a saturation when the strength of the
EGMF is $\sim 10^{-11}$ G, it is very possible that the
contribution from the cascade emission produced by blazars can be
considered as another method to probe the upper limit of the
strength of the EGMF.

\end{abstract}

\begin{keywords}
diffuse radiation -- galaxies: active -- gamma rays: general
\end{keywords}

\section{Introduction}

The gamma-ray sky observed by the Large Area Telescope (LAT) on
board the {\it Fermi Gamma Ray Space Telescope} consists of the
resolved gamma-ray emitters, such as normal galaxies, active
galactic nuclei (AGNs), gamma-ray bursts (GRBs), and pulsars etc.,
and the diffuse gamma-ray radiation including emission from the
Galaxy and the extragalactic gamma-ray background (EGRB)
\citep[e.g.,][]{abdo10a,abdo10b}. The origin of the EGRB is poor
known, which is one of the fundamental unsolved problems in
astrophysics. The EGRB was first detected by the SAS-2 satellite
\citep{Fichtel75}. Latter, in the 1990s the Energetic Gamma-Ray
Experiment Telescope (EGRB) on board the {\it Compton Observatory}
measuerd its spectrum at 0.03 -- 50 GeV with a good accuracy
\citep{Sreekumar98,Strong04}. Recently, LAT has made a new
measurement of the EGRB spectrum \citep{abdo10b}. This has been
found to be consistent with a featureless power-law with a photon
index of $\sim$2.41 in the 0.2 -- 100 GeV energy range. The
observed integrated flux above 100\ MeV is $(1.03\pm0.17) \times
10^{-5}\ {\rm photons}\ {\rm cm^{-2}}\ {\rm s^{-1}}\ {\rm
sr^{-1}}$, which is lower than the one of $(1.14\pm0.05) \times
10^{-5}\ {\rm photons}\ {\rm cm^{-2}}\ {\rm s^{-1}}\ {\rm
sr^{-1}}$ observed by EGRET \citep{Strong04,abdo10b}.

Various kinds of unresolved gamma-ray sources have been proposed
as the possible contributors to the EGRB, such as AGNs
\citep[e.g.,][]{Padovani93,Chiang98,Giommi06,Narumoto06,Dermer07,Cao08,Inoue09,Venters09,Venters10,Inoue11,Ajello11},
star-forming galaxies
\citep[e.g.,][]{Pavlidou02,Fields2010,Makiya2011}, and starburst
galaxies \citep[e.g.,][]{Thompson07,Makiya2011}. Since blazars are
the dominant extragalactic gamma-ray sources, it is naturally
expected that the intrinsic emission of an unresolved population
of blazars would account for a sizable contribution to the EGRB.
However, because of the uncertainties of gamma-ray luminosity
function (GLF) and the spectrum index distribution (SID) of
blazars, this contribution to the EGRB from blazars has been
debated. \citet{Stecker11} and \citet{Inoue09} suggested that this
contribution is dominated by the emission from unresolved blazars,
while \citet{Ajello11} and \citet{abdo10c} argued that unresolved
blazars produce a small fraction of the EGRB.

Additionally, very high energy (VHE) photons from blazars will
interact with the UV-IR photons of extragalactic background light
(EBL) to produce pairs of electrons and positrons, and the pairs
will boost cosmic microwave background (CMB) photons to gamma-ray
energies through inverse Compton scattering process
\citep[e.g.,][]{Dai2002,Yang2008,Tavecchio10,Neronov2010}. Cascade
contributions to the EGRB have been considered by many authors
\citep[e.g.,][]{Strong73,Strong74,Kachelriess}. The cascade flux
produced by VHE photons of blazars, which is lower than the LAT
sensitivity, will make contribution to the EGRB. Besides, for a
large extragalactic magnetic field (EGMF) strength, the cascade
photons will spread over a large angle, so that LAT cannot
identify this emission from a point source, which gives rise to
another contribution to the EGRB. The latter contribution depends
on the strength of the EGMF. In the case of high collective
high-energy intensities of blazars or high EBL intensities, the
cascade contribution to the EGRB is significant \citep{Coppi97}.
\citet{Kneiske08} have estimated the contribution radiation from
the first generation pairs to the EGRB, and they have found that
this cascade contribution is sizable. But \citet{Inoue09}
suggested that this cascade contribution is negligible. Very
recently, \citet{Venters10} has also estimated this cascade
contribution by using a Monte Carlo program, and found that the
cascade radiation greatly enhances the contribution to the EGRB
from blazars. It should be noted that \citet{Kneiske08} and
\cite{Inoue09} have taken neither the SID of blazars nor the EGMF
into account, \citet{Venters10} included the effect of the SID of
blazars but ignored the effect of the EGMF on the cascade
contribution.

In this paper, we revisit the cascade contribution to the EGRB
from blazars by assuming that the observed LAT GeV spectrum could
be extrapolated into TeV energy range with a power-law spectrum
and the SID, in particular, we study the impact of the EGMF on the
spectrum of the cascade radiation. In \S 2, we present the method
we used to calculate the contribution from blazars and present our
results.  Finally  we give a brief discussion and conclusions in
\S 3.

\section{Contribution of the EGRB from blazars}

\subsection{Blazar gamma ray luminosity function}

To estimate the contribution to the EGRB from unresolved blazars,
The blazar GLF is required. However, due to the small sample size,
it is difficult to build a GLF directly using the current
gamma-ray-loud blazar sample. On the other hand, luminosity
functions (LFs) of blazars at other wavelengths (like radio or
X-ray) are widely studied in previous works
\citep[e.g.,][]{Dunlop1990,Hasinger2005}. Moreover, it is well
believed that the gamma-ray emission of blazars would correlate
with the emissions at lower energy bands. For instance, it was
found that there is a good correlation between the gamma-ray
luminosity and radio luminosity \citep[e.g.,][]{Ackermann11,
Ghirlanda11}, and \citet{Ghisellini10} found a positive
correlation between the jet power and the luminosity of the
accretion disc in some blazars. Hence, the GLF is typically built
from the luminosity functions in other wavelengths by using the
correlation between gamma-ray emission and lower energy band
emissions.

For the purpose of our analysis, we will adopt the best-fit models
of blazar GLF constructed by \citet{Narumoto06}: pure luminosity
evolution (PLE) and luminosity-dependent density evolution (LDDE)
models. \citet{Narumoto06} limited the model parameters by using
likelihood analysis of the observed redshift and gamma-ray flux
distributions of the EGRET blazars and found that the LDDE model
gives a better fit to the observed distributions than the PLE
model. For the blazar GLF $\rho_\gamma(L_\gamma, z)$ and its
parameters in each model, we totally use them given by
\citet{Narumoto06}.

\subsection{The intensity of the EGRB from intrinsic emission of blazars }

In order to estimate the EGRB intensity from the intrinsic
emission of  blazars, we need the SID of Fermi-LAT resolved
blazars. In the clean sample of the First LAT AGN Catalog (1LAC),
corresponding to 11 months of data collected in science operation
mode, there are 523 blazars \citep{abdo10a}. Their energy spectra
between 0.1 GeV to 100 GeV can be fitted with a power-law spectral
with photon index $\Gamma$. Their photon index distribution is
compatible with a Gaussian distribution. However, as argued by
\citet{abdo10a}, sources with hard spectra are more easily
detected by LAT, therefore the observed photon index distribution
tends to harder than the intrinsic one. As shown in
\citet{abdo10a} (see their Fig.1), when the integrated flux
$F_{\gamma,>100}$ above 100 MeV is larger than $7 \times 10^{-8}\
{\rm photon}\ {\rm cm^{-2}}\ {\rm s^{-1}}$, LAT detected all
sources irrespective of their photon indices, fluxes, or positions
in the specific sky. For the sources with $F_{\gamma,>100}\geq7
\times 10^{-8}\ {\rm photon}\ {\rm cm^{-2}}\ {\rm s^{-1}}$, their
spectral photon index distribution, which is the intrinsic SID, is
also compatible with a Gaussian distribution and is given by
\citep{abdo10a}
\begin{equation}
\frac{dN}{d\Gamma}=e^{-\frac{(\Gamma-\mu)^2}{2\sigma^2}}
\end{equation}
with a mean of $\mu=2.40\pm0.02$ and a dispersion of
$\sigma=0.24\pm0.02$ .

\begin{figure}
\begin{center}
 \includegraphics[width=90mm]{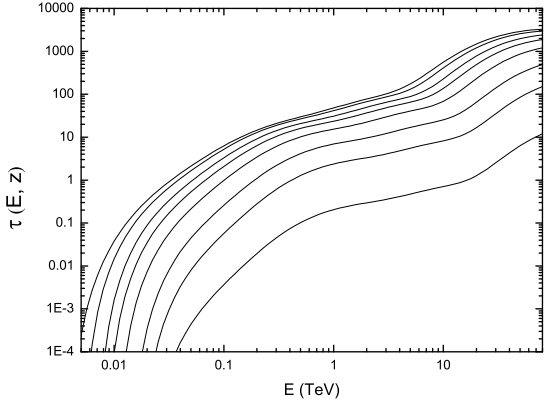}
 \caption{The optical depths expected in the EBL model of \citet{Finke2010} as a function of the
photon energy $E$ for sources located at $z=$0.02, 0.2, 0.5, 1.0,
1.5, 2.0, 3.0, 4.5 from bottom to top. } \label{fig1}
\end{center}
\end{figure}

When the GLF and the intrinsic SID  are given, the EGRB intensity,
$F^{\rm i}_{\rm EGRB}(E)$, in units of photons\ ${\rm cm^{-2}}\
{\rm s^{-1}}\ {\rm sr^{-1}}\ {\rm MeV^{-1}}$, from intrinsic
emission of blazars is given by
\begin{eqnarray}
F^{\rm i}_{\rm EGRB}(E)&=&\eta\int_{\Gamma_{\rm min}}^{\Gamma_{\rm
max}}d\Gamma \frac{dN}{d\Gamma} \int_{0}^{z_{\rm
max}}dz\frac{d^2V}{dzd\Omega}\\\nonumber
 & &\int^{L_{\gamma, \rm max}}_{L_{\gamma,
 \rm min}}dL_{\gamma}\rho_{\gamma}F^{\rm
 i}_{\gamma}(E,z,L_{\gamma},\Gamma)\\\nonumber
& & \times e^{-\tau(E,z)}(1-\omega(F_{\gamma,>100}))\;\ ,
\label{Fdir}
\end{eqnarray}
where $\Gamma_{\rm min}=1.2$ and $\Gamma_{\rm max}=3.0$ are
minimum and maximum values of the photon index, $z_{\rm max}=5.0$
is the maximum redshift, $L_{\rm \gamma, min}=10^{44}\ {\rm erg}\
{\rm s^{-1}}$ and $L_{\rm \gamma, max}=10^{52}\ {\rm erg}\ {\rm
s^{-1}}$ are the minimum and maximum luminosities respectively;
$F^{\rm i}_{\gamma}(E,z,L_{\gamma},\Gamma)$ is the intrinsic
photon flux at energy $E$ of a blazar with gamma-ray luminosity
$L_{\gamma}$ and a power-law spectrum at redshift $z$ and is given
by \citep{Venters09}
\begin{equation}
F^{\rm i}_{\gamma}(E,z,L_{\gamma},\Gamma)=\frac{L_{\gamma}}{4\pi
d^2(z)\cdot E^2_0}(1+z)^{2-\Gamma}(\frac{E}{100\ \rm
MeV})^{-\Gamma}\;,
\end{equation}
where $E^2_0=100\  {\rm MeV}\cdot 1.6\times10^{-4}\ {\rm erg}$,
$\tau(E,z)$ is the optical depth of the EBL for the sources at
redshift $z$ emitting gamma-ray photon energy $E$, here we use the
model of \citet{Finke2010} to derive $\tau(E,z)$; The optical
depths expected in this EBL model as a function of the photon
energy $E$ for sources at different redshifts are presented in
Fig.\ref{fig1}. $\omega(F_{\gamma,>100})$ is the detection
efficiency of LAT at the photon flux $F_{\gamma,>100}$ , which
corresponds to the integrated flux above 100 MeV from the sources
at redshift $z$ with $L_{\gamma}$ and $\Gamma$, and
$F_{\gamma,>100}=F_{\gamma}(100\ {\rm
MeV},z,L_{\gamma},\Gamma)\times100\ {\rm MeV}/(\Gamma-1)$
\citep{Venters09}; $\eta$ is the normalization factor and
$\eta=\frac{N_{\rm obs}}{N_{\rm exp}}$, where $N_{\rm obs}$ is the
number of the sample, and $N_{\rm exp}$ is the expected number and is
given by
\begin{eqnarray}
N_{\rm exp}&=&4\pi \int_{\Gamma_{\rm min}}^{\Gamma_{\rm
max}}d\Gamma \frac{dN}{d\Gamma} \int_{0}^{z_{\rm
max}}dz\frac{d^2V}{dzd\Omega}\\\nonumber
& & \int^{L_{\gamma, \rm
max}}_{L_{\gamma,
 \rm min}}dL_{\gamma}\rho_{\gamma}\omega(F_{\gamma,>100})\;\ .
\end{eqnarray}
Here, the detection efficiency of LAT reported by \citet{abdo10a}
and $N_{\rm obs}=523$ are used.

\subsection{The intensity of the EGRB from cascade emission of blazars  }

For the cascade emission of blazars, we follow the geometry of the
cascade process given by \cite{Dermer11}. In this geometry, VHE
photons emitted at an angle $\theta_1$  by a blazar at distance
$d$ could convert into electron-positron pairs via photon-photon
absorption with the photons of the EBL after travelling a mean
distance $\lambda_{\gamma\gamma}=d/\tau(E_{\rm VHE},z)$. When the
pairs are deflected by the EGMF by an angle $\theta_{\rm dfl}$,
these pairs could scatter CMB photons to gamma-ray energies, where
$\theta_{\rm dfl}=\lambda_{\rm T}/r_{\rm L}$ for $\lambda_{\rm
T}<\lambda_{\rm coh}$, and $\theta_{\rm dfl}=\lambda_{\rm
T}/r_{\rm L}(\lambda_{\rm coh}/\lambda_{\rm T})^{1/2}$ for
$\lambda_{\rm T}>\lambda_{\rm coh}$, $\lambda_{\rm T}$ is  the
cooling length for electrons with energy $\gamma$ due to Thomson
scattering, $r_{\rm L}$ is the electron/positron Larmor radius,
and $\lambda_{\rm coh}$ is the coherence length of the EGMF
\citep{Dermer11}. The redirected secondary gamma-ray photons would
arrive at an angle $\theta=\lambda_{\gamma\gamma}\theta_{\rm
dfl}/d$ to the line of sight. In calculations, $\lambda_{\rm
T}=3m_{e}c^2/4\sigma_{\rm T}U_{\rm CMB}(1+z)^4\gamma$, where
$\sigma_{\rm T}$ is the Thomson cross section, and $m_{\rm e}$ is
the rest mass of electron, and $U_{\rm CMB}=4\times 10^{-13}\ {\rm
erg}\ {\rm cm^{-3}}$ is the CMB energy denstiy at $z=0.0$; $r_{\rm
L}=m_{\rm e}c^2\gamma/eB_{\rm EGMF}$, where $B_{\rm EGMF}$ is the
EGMF of strength and $e$ is the elementary charge. Here, we set
the coherence length of the EGMF $\lambda_{\rm coh}=1$\ Mpc.

There are two possible cases for the contributions of the cascade
emission to the EGRB.  In the first case (case I), the cascade
emission could make contribution to the EGRB if the flux of the
cascade emission is lower than the LAT sensitivity. In the second
case, although the flux of the cascade emission is larger than the
LAT sensitivity, the angle between of the redirected secondary
gamma-ray photons and the line of sight is larger than LAT
point-spread-function (PSF) angle, i.e. $\theta>\theta_{\rm PSF}$,
the cascade emission also will not be attributed to a point source
by LAT, and then make contribution to the EGRB, where $\theta_{\rm
PSF}=\frac{1.7\pi}{180}(0.001E)^{-0.74}[1+(0.001E/15)^2]^{0.37}$
\citep{Taylor11}. In the following, we will calculate the
contributions of cases I and II, respectively.

To calculate the cascade flux, the intrinsic TeV flux is
necessary, which is poor known due to the uncertainties of physics
properties of blazars and the EBL. An approach to derive the
possible intrinsic TeV spectrum is to extrapolate the observed LAT
GeV spectrum to TeV energy range by using a power-law spectrum.
Here, we consider the extrapolated TeV spectrum as the intrinsic
TeV spectrum, which is written as
\begin{equation}
F_{\rm VHE}^{\rm intrinsic}(E_{\rm
VHE},z,L_{\gamma},\Gamma)=\frac{L_{\gamma}(1+z)^{2-\Gamma}}{4\pi
d^2(z)E^2_0}(\frac{E_{\rm VHE}}{100\ \rm MeV})^{-\Gamma}\;\ .
\end{equation}
Therefore, the total cascade flux of inverse Compton-scattering
CMB photons produced by the first generation of electrons through
photon-photon pair production is given as \citep{Dermer11,Huan11}
\begin{eqnarray}
F^{\rm
cascade}_{\gamma}(E,z,L_{\gamma},\Gamma)=\frac{81\pi}{16\lambda_{\rm
c}^3}\frac{\epsilon_{\rm c}^2 m_{\rm e}c^2}{(1+z)^4U_{\rm
cmb}}\\\nonumber \int_{\sqrt{\frac{3\epsilon_{\rm
c}}{4\epsilon_{\rm CMB}(1+z)}}}^{\infty}\frac{d\gamma}{\gamma^8
{\rm exp}(3\epsilon_{\rm c}/4\gamma^2 \epsilon_{\rm CMB}(1+z))-1}
\\\nonumber
 \times \int_{2\gamma}^{\epsilon_{\rm max}}d\epsilon F_{\rm
VHE}^{\rm
intrinsic}(\epsilon\frac{5.11\times10^5}{10^6},z,L_{\gamma},\Gamma)(1-e^{-\tau(\epsilon,z)})\;
,
\end{eqnarray}
where the dimensionless energy $\epsilon_{\rm c}=E\cdot
10^6/(5.11\cdot10^5)$, and $\epsilon_{\rm CMB}=1.24\times10^{-9}$
in $m_{\rm e}c^2$ units is the average CMB photon energy at
$z=0.0$, and $\lambda_{\rm c}=2.426\times10^{-10}$\ cm is the
Compton length. For the maximum dimensionless energy of intrinsic
TeV photons $\epsilon_{\rm max}$, we take $\epsilon_{\rm
max}=2.0\times10^8$, corresponding to  $E_{\rm VHE}=100\ $TeV. The
emission from the secondary generation electrons or more than
secondary generation electrons is negligible at GeV range
\citep{Kneiske08}. The energy losses of the electrons by
synchrotron emission are negligible since the ratio between the
cooling times of inverse Compton scattering CMB photons and
synchrotron emission is very small, $\sim10^{-6}$ even when the
upper limit value of the EGMF $\sim10^{-9}$ G is used.

The EGRB intensity in case I, independent of the EGMF, is
given by
\begin{eqnarray}
F^{\rm c,i}_{\rm EGRB}(E)&=&\eta\int_{\Gamma_{\rm
min}}^{\Gamma_{\rm max}}d\Gamma \frac{dN}{d\Gamma}
\int_{0}^{z_{\rm max}}dz\frac{d^2V}{dzd\Omega}\\\nonumber &
&\int^{L_{\gamma, \rm max}}_{L_{\gamma,
 \rm min}}dL_{\gamma}\rho_{\gamma}F_{\gamma}^{\rm cascade}(E,z,L_{\gamma},\Gamma)
\\\nonumber
 & &\times e^{-\tau(E,z)}(1-\omega(F^{\rm cascade}_{\gamma,>100}))\;\ ,
\end{eqnarray}
where $F^{\rm cascade}_{\gamma,>100}=2F^{\rm
cascade}_{\gamma}(E,z,L_{\gamma},\Gamma)100\ {\rm MeV}$
$\cdot(100\ {\rm MeV}/E)^{-1.5}$, which is derived by assuming
that the cascade spectrum is a power-law spectrum with photon
index 1.5 \citep{Strong74,Tavecchio10}.

The EGRB intensity in case II, which depends on the EGMF, is
written as,
\begin{eqnarray}
F^{\rm c,ii}_{\rm EGRB}(E)&=&\eta\int_{\Gamma_{\rm
min}}^{\Gamma_{\rm max}}d\Gamma \frac{dN}{d\Gamma}
\int_{0}^{z_{\rm max}}dz\frac{d^2V}{dzd\Omega}\\\nonumber
&&\int^{L_{\gamma, \rm max}}_{L_{\gamma,
 \rm min}}dL_{\gamma}\rho_{\gamma}F_{\gamma,\theta>\theta_{\rm PSF}}^{\rm cascade}(E,z,L_{\gamma},\Gamma)
\\\nonumber
 &&\times e^{-\tau(E,z)}\omega(F^{\rm cascade}_{\gamma,>100})\;\ .
\end{eqnarray}
In case II, to avoid the situation that the produced pair are
isotropized by EGMF when the EGMF of strength is large
enough, we set $\theta_{\rm dfl}<\pi/2$.

\subsection{Total EGRB intensity of blazars and its contribution to the EGRB}

As mentioned above, there are two possible cases (cases I and II)
for the contributions of the cascade emission to the EGRB. If we
only consider case I for the cascade emission, i.e., the flux of
the cascade emission is lower than the LAT sensitivity, the total
EGRB intensity of blazars is given by
\begin{equation}
F^{\rm tot}_{\rm EGRB}(E)=F^{\rm i}_{\rm EGRB}(E)+F^{\rm c,i}_{\rm
EGRB}(E)\;.
\end{equation}
The cascade emission is independent on the EGMF in this case. In
fact, the contribution of the cascade emission to the EGRB would
include the contributions in both cases I and II (in the latter
the flux of the cascade emission is larger than the LAT
sensitivity, however the angle between the redirected secondary
gamma-ray photons and the line of sight is larger than LAT PSF
angle). Therefore, the total EGRB intensity of blazars is given by
\begin{equation}
F^{\rm tot}_{\rm EGRB}(E)=F^{\rm i}_{\rm EGRB}(E)+[F^{\rm
c,i}_{\rm EGRB}(E)+F^{\rm c,ii}_{\rm EGRB}(E)]\;.
\end{equation}
It should be noted that the cascade emission depends on the EGMF
in this case.

\begin{figure}
\begin{center}
 \includegraphics[width=90mm]{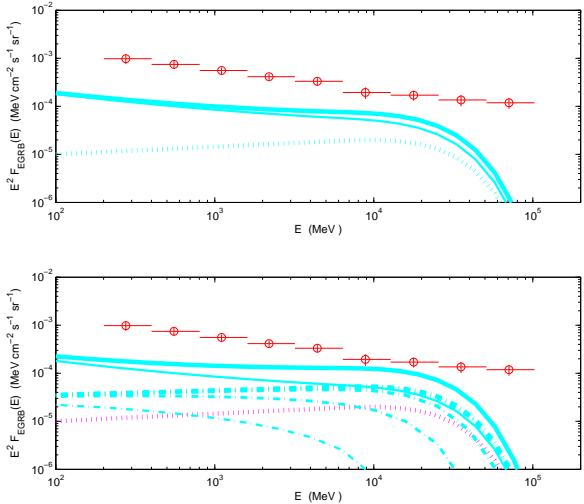}
\caption{Blazar contribution to the EGRB assuming the LDDE model
of the blazar GLF. Top: the calculation of the blazar intrinsic
contribution (thin solid line) and the cascade contribution of
case I (dotted). The thick solid line is the sum of the two
contributions. Bottom: comparison of the blazar intrinsic
contribution (thin solid line) and the cascade contributions in
case I (dotted line) and in case II (dot-dashed line ). For the
cascade contribution of case II, we calculate with $B_{\rm
EGMF}=10^{-14}, 10^{-13}, 10^{-12}, 10^{-11}, 10^{-10}$\ G
(dot-dashe line: from thin to thick), respectively. The thick
solid line is the sum of the blazar intrinsic contribution and the
cascade contributions of cases I and II with $B_{\rm
EGMF}10^{-11}\ $G. } \label{fig2}
\end{center}
\end{figure}

At first we calculate the intrinsic emission, the cascade
emission, and the total EGRB intensities for the LDDE model of the
blazar GLF and the calculated results are shown in Fig.\ref{fig2}.
In the top panel of Fig.\ref{fig2}, we consider the case I of the
blazar cascade emission, i.e., the total intensity is given by Eq.
(9) and the cascade emission is independent on the EGMF; thin
solid, dotted, and thick solid lines represent the intrinsic,
cascade, and total intensities, respectively. We have found that
the contribution of the blazar cascade emission in case I to the
EGRB could not be negligible, and would enhance the contribution
of the blazar emission to the EGRB at high energies. At lower
energies, however, this contribution is negligible.

We now consider the contributions of the cascade emission in both
cases I and II, i.e., the total intensity is given by Eq. (10).
Since the cascade emission in case II depends on the EGMF, we
calculate the intensities with different strengthes of the EGMF.
The results are shown in the lower panel of Fig.\ref{fig2}. It can
be seen from the figure that the intensity of the blazar cascade
emission in case II increases with increasing of the strength of
the EGMF, and will reach a saturation at some value of the EGMF.
From our calculations, the contribution of the blazar cascade
emission in case II reaches a saturation when the strength of the
EGMF is $B_{\rm EGMF}\approx 10^{-11}\ $G; the contribution of the
blazar cascade emission with $B_{\rm EGMF}<10^{-14}\ $G in case II
is negligible, and this contribution becomes significant compared
with the contribution of the blazar intrinsic emission at several
GeV to several tens GeV with $B_{\rm EGMF}>10^{-13}\ $G. If the
strength of the EGMF satisfies that $B_{\rm EGMF}>10^{-12}\ $G,
the total contribution of blazars can account for the large
fraction of the EGRB at several GeV to several tens GeV.

\begin{figure}
\begin{center}
 \includegraphics[width=90mm]{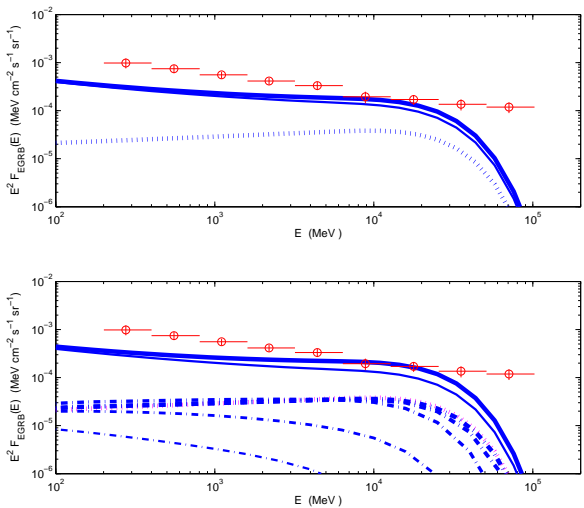}
 \caption{Same as Fig.\ref{fig2} except for the PLE model of the blazar GLF. }
\label{fig3}
\end{center}
\end{figure}

In Fig.\ref{fig3}, we present the results by using the PLE model
of the blazar GLF. It is found that the contributions of both the
blazar intrinsic emission and the cascade emission in case I are
significantly larger than those derived by using the LDDE model of
the blazar GLF. The features of the contribution from the blazar
cascade emission in case II is similar with those in Fig.
\ref{fig2}. If the strength of the EGMF is larger ( e.g., $B_{\rm
EGMF}>10^{-12}\ $G), the contribution of the cascade emission  to
the EGRB is also significant. Assuming the PLF model, the total
contribution of blazars can explain the bulk of the EGRB,
irrespective of the strength of the EGMF.  Moreover, it seems that
the model of the blazar GLF has an impact on the contributions to
the EGRB for both the cascade emission in case I and the blazar
intrinsic emission.

\section{DISCUSSION AND CONCLUSIONS}

We have studied the impact  of the cascade radiation on the
contribution of blazars to the EGRB by using  a simple
semi-analytic model. In particular, we take the effect of the EGMF
on the cascade contribution from blazars into account. We have
found that if the strength of the EGMF is large enough ($B_{\rm
EGMF}>10^{-12}$\ G), the cascade contribution could significantly
alter the spectrum of the EGRB at high energies, while with small
strength of the EGMF large enough ($B_{\rm EGMF}<10^{-14}$\ G),
the cascade contribution is small, but cannot be ignored (see
Figs. 2 and 3).

There is little information about the strength of the EGMF up to
now. Measurements of Faraday rotation of polarization of radio
radiation from distant quasars place upper limits on the EGMF
strengths at the level of $\sim10^{-9}\ $G
\citep[e.g.,][]{Kronberg94,Blasi99}. On the other hand, the
possible cascade emission from resolved blazars provides a method
to probe the lower bands on the EGMF. The lower bounds on the EGMF
at the level of $10^{-18}\ $ -$10^{-15}\ $G were obtained by using
this method
\citep[e.g.,][]{Taylor11,Dermer11,Huan11,Tavecchio10,Neronov2010,Dolag11}.
In this paper, we have shown that the contribution of the blazar
cascade emission to the EGRB very possibly provides another method
to set the upper bands of the EGMF. In order to do so, the
prerequisites are that we have the well determined GLF and we can
well estimate the contributions of other sources to the EGRB. We
have shown that a possible upper bound of the EGRB is $\sim
10^{-11}$ G.

In estimating the cascade contribution to the EGRB, there is an
uncertainty in the EBL model. \citet{Venters10} have found that
the amount of cascade radiation is sensitive to the EBL model.
Actually, one of the EBL models \citet{Venters10} adopted is high
level \citep{Stecker06}, and the other one is low level
\citep{Gilmore09}. However, recent study indicated that the EBL
intensity is low \citep{abdo10d}, and the commonly used several
EBL models are low levels and have similar results
\citep[e.g.,][]{Finke2010,Dom11}. It is therefore expected that
the impact of uncertainty of the EBL model on the contribution of
the cascade emission is slight.

Based on the observed gamma-ray sample, various models of the
blazar GLF have been constructed
\citep[e.g.,][]{Stecker11,Ajello11,Inoue09}. \citet{Ajello11} have
found that these models at redshift zero are relative consistent,
while the significant discrepancy between these models will appear
at redshift one. The LDDE model we used here expects smaller
number of  blazars compared with those expected in the models of
\citet{Stecker11}, \cite{Ajello11}, and \cite{Inoue09}. The model
of \citet{Stecker11} is similar with the PLE model we used in this
paper. Actually, the determination of the blazar GLF solely based
on the observed gamma-ray sample is not certain, due to the
sources confusion effect and the active feature of blazars
\citep{Venters11}. \citet{Stecker11} indicated that the sources
confusion would lead to the underestimate of the contribution to
the EGRB below 1 GeV from blazars. The active flare can bring the
intrinsically faint blazars to the bright blazars sample, which
will flatten the faint-end slope of the observed blazar source
counts. This flare effect also can cause underestimate of the
contribution. In this sense, we conclude that we cannot rule out
blazar dominance of the EGRB from our calculations.

\section*{Acknowledgments}
We thank the anonymous referee for his/her very constructive
comments to substantially improve the quality of this paper. This
work is partially supported by the National Natural Science
Foundation of China (NSFC 10778702), a 973 Program (2009CB824800),
and Yunnan Province under a grant 2009 OC.


\bibliography{refernces}


\end{document}